\documentclass[12pt]{article}
\usepackage{graphicx}
\usepackage{amsmath}
\usepackage[cp866]{inputenc}
\usepackage[english]{babel}
\usepackage{xcolor}
\usepackage{epsfig}
\usepackage{floatrow}
\usepackage{amssymb,amsmath,amsfonts,amsthm,graphicx,psfrag}
\usepackage{indentfirst}
\usepackage{hyperref}
\usepackage[title,titletoc]{appendix}
\usepackage{graphicx}
\usepackage{amsfonts}
\usepackage{bm}
\usepackage{verbatim}
\usepackage{epsfig}
\graphicspath{{./pictures/}}
\usepackage{authblk}
\usepackage{cite}
\usepackage[title,titletoc]{appendix}
\usepackage{slashed}
\usepackage{amsmath}
\usepackage{physics}

		\newcommand{\be}{\begin{eqnarray}}
\newcommand{\ee}{\end{eqnarray}}

\newcommand{\ba}{\begin{aligned}}
\newcommand{\ea}{\end{aligned}}

\title{The spatial string tension  from the Field Correlator Method}
\author{ N. O. Agasian$^{+}$, Z. V. Khaidukov$^{+,a}$\footnote{e-mail:khaidukov.zv@phystech.edu} and  Yu. A. Simonov$^{+}$  \\
$^{+}$NRC ``Kurchatov Institute'', Moscow, Russia \\
$^{a}$ Moscow Institute of Physics and Technology, 9 Institutskiy per., Dolgoprudny, Moscow region, 141701, Russia}

\newcommand{\beq}{\begin{eqnarray}}
 \newcommand{\eeq}{\end{eqnarray}}

\def\fun#1#2{\lower3.6pt\vbox{\baselineskip0pt\lineskip.9pt
\ialign{$\mathsurround=0pt#1\hfil ##\hfil$\crcr#2\crcr\sim\crcr}}}

\newcommand{{\SD}}{\rm SD}

\newcommand{{\Mc}}{\mathcal{M}}

\newcommand{\ver}{\mbox{\boldmath${\rm r}$}}

\newcommand{\vep}{\mbox{\boldmath${\rm p}$}}

\newcommand{\vez}{\mbox{\boldmath${\rm z}$}}

\newcommand{\lan}{\langle}
\newcommand{\ran}{\rangle}
\renewcommand{\vec}[1]{\boldsymbol{#1}}

\begin{document}
\maketitle
\begin{abstract}

The behaviour of  spatial string tension $\sigma_s(T)$ as a function of the temperature $T$ is found in the framework of the Field Correlator Method
(FCM). Here the string tension is calculated using the gluelump Green's function, where gluons in the gluelump are interacting via the same spatial string tension. The resulting  T-dependence was obtained without extra parameters in the region
$T_c < T < 5 T_c$ using the formalism of elliptic  functions $\theta_3$,
demonstrating  good agreement with available lattice data.

\end{abstract}

\section{Introduction}
The confinement in QCD is a basic phenomenon which ensures more than 90 percent of the visible mass in the Universe and makes the world such as we see it. At zero temperature
the theory of confinement in QCD was formulated in the framework of the Field Correlator Method (FCM) \cite{1,2,3,4,5,5*},
 via the vacuum field correlators  of the
colorelectric (CE) and the colormagnetic (CM) fields $E_i^a,H_i^a$, and at the temperature
$T=0$ the behaviour of all physical quantities is expressed  via the basic nonperturbative
parameter -- the string tension, which  can have different values in the light-like  $\sigma_E$ and space-like $\sigma_H$ areas, but  at  zero temperature T,  $\sigma_E(T=0)=\sigma_H(T=0)= \sigma$.  Very important role in FCM plays
bilocal correlator (BC) of gluonic fields strength:
	\be
	\frac{g^{2}}{N_{c}}<tr_{f}\Phi(y,x)F_{\mu\nu}(x)\Phi(x,y)F_{\lambda \rho}(y)>\equiv D_{\mu\nu,\lambda\rho}(x,y) \label{eqcorr}.
	\ee
From this moment we use     $F_{\mu\nu}\equiv F^{a}_{\mu\nu}T^{a}$, $a=1..N_{f},T^{a}$ - are generators of fundamental representation of  $SU(N_{c})$
. In Eq. (\ref{eqcorr}) symbol $<>$  means  averaging over Yang-Mills action  $S=\frac{1}{4g^{2}}\int{d^{4}x}(F^{a}_{\mu\nu}), F^{a}_{\mu\nu}=\partial_{\mu}A_{\nu}-\partial_{\nu}A_{\mu}+gf^{abc}A^{b}_{\mu}A^{c}_{\nu}, a=1..N^{2}_{c}-1$, $\Phi(x,y)=Pexp(i\int^{x}_{y}A_{\mu}dz^{\mu}), \mu=1..4$\footnote{We work in Euclidean space, the fourth component plays role of Euclidean time.} is  the Wilson line in fundamental representation.
	 We  can write BC as follows:
	\be
	 D_{\mu\nu,\lambda\rho}(x,y)=(\delta_{\mu\lambda}\delta_{\nu\rho}-\delta_{\mu\rho}\delta_{\nu\lambda})D(x-y)+\frac{1}{2}(\frac{\partial}{\partial x_{\mu}}(x-y)_{\lambda}\delta_{\nu\rho}+perm.))D_{1}(x-y)  \label{eqcorr1},
	\ee
	$D(x-y),D_{1}(x-y)$ - are scalar functions.  We also can add index E or H\footnote{We will write them only where we need to avoid ambiguity.} to  $D,D_{1}$, because $E_{i}=F_{0i},H_{i}=\epsilon_{ijk}F^{jk}/2 , i=1,2,3$ and $<EH>=0$.
	Functions  $D^{E,H}(x),D^{E,H}_{1}(x)$   define all confining QCD dynamics and in particular the string tensions:
\be
\sigma_E= 1/2 \int (d^2z)_{i4} D^{E}(z), \sigma_H= 1/2 \int (d^2z)_{ik} D^{H}(z) \label{2}. \ee
These functions were calculated in good agreement
between the FCM \cite{6} and the lattice data \cite{3,5}, while $D^E(x),D^H(x)$ were also studied
in details at $T>0$ on the lattice \cite{5}. The most interesting fact is  that
 at $T>0$  $\sigma_E(T)$ and $\sigma_H(T)= \sigma_s(T)$
behave differently. Namely:  $\sigma_E(T)$ displays a spectacular drop before $T=T_c$ and
disappears above $T=T_c$, while in contrast to that $\sigma_s(T)$
grows almost quadratically at large $T$, as  was
found on the lattice \cite{4,7*,8,9} and supported by the studies
in the framework of the FCM \cite{10,10*,11,12,12*,12**,18,15,14}.
Indeed, as it was found in Ref.\cite{8,9} that the dominant part of the spatial string tension $\sigma_s(T)$ grows quadratically at large $T$
\be
\sigma_s(T)= (c_{\sigma})^2 g^4(T) T^2  \label{3}, \ee
where $c_{\sigma}$ was defined numerically in the lattice calculations \cite{8,9} in  case  $N_c=3, N_f=0$ as
\be
c_{\sigma}= 0.566\pm 0.013.  \label{4} \ee

 On the theoretical side the quadratic growth of the $\sigma_s(T)$ was derived in the framework of FCM
 \cite{6,10,11,12}, and the  value of $c_{\sigma}$ was found in Ref.  \cite{12*}
 in  good agreement with the lattice data of \cite{8,9}.

However, in the full FCM expression for the spatial string tension
the term in the  Eq. (\ref{3}) is only a fast-growing part  of the whole expression, which was hitherto not known.

The purpose of this paper is to derive the total expression of the spatial string tension including the linear in $T$ part, to calculate the numerical value of $\sigma_s(T)$  in the  region of $[T_{c}..5T_{c}]$\footnote{In FCM $\sigma_{s}$ below $T_{c}$ almost coincides with the  vacuum  value $\sigma$.} and compare it with the lattice data.
As will be seen, the results, obtained within the FCM,  will provide the values
of $\sigma_s (T)$ in  good agreement with the lattice data. In the next section we discuss connection  between BC and gluelump Green's function.
In section 3 we discuss the general expression for $\sigma_s(T)$ in terms of the field correlators (and gluelump Green's functions respectively), we  formulate
 its final form and  obtain  the expression of the string tension, which is discussed and compared with the lattice data in  section 4. The renormalized  coupling constant $g^2(T)$ is  given in Appendix A1, the detailed discussion of two-gluelump Green's function  is given in Appendix A2. We discuss diagonalization  of the  two-glulump Hamiltonian   in Appendix A3. Calculations of the two-gluelump Hamiltonian eigenvalues and eigenfunctions is given   in Appendix A4.
	\section{Gluelump and bilocal correlator of gluonic field strength}
 We need to find a connection between
$D(x)$ and so called  gluelump Green's function \cite{11*}. For this purpose we rewrite the expression in Eq. (\ref{eqcorr1}) in the  form:
\be
	\frac{g^{2}}{N_{c}}<tr_{f}\Phi(y,x)F_{\mu\nu}(x)\Phi(x,y)F_{\lambda \rho}(y)>=	 \frac{g^{2}}{N_{c}}tr_{f}<F^{a}_{\mu\nu}(x)[T^{a}\Phi(x,y)T^{b}\Phi(y,x]F_{\lambda \rho}(y)>.
\ee
The integration  in the last expression  is
performed along the straight line connecting the
points x and y, so we can rewrite  the expression in square brackets in the form \cite{10*}:
\be
2tr(T^{a}\Phi(x,y)T^{b}\Phi(y,x))=\Phi^{ab}_{adj}(x,y),
\ee
and finally we have:
\be
D_{\mu\nu,\lambda\rho}(x,y)=\frac{g^{2}}{2N^{2}_{c}}tr_{adj}<F^{a}_{\mu\nu}(x)\Phi^{ab}_{adj}(x,y)F^{b}_{\lambda\rho}(y)>.
\ee
Last equation  coincides with the expression from the paper \cite{Dosh}. For our purposes we  rewrite it as:
\be
D_{\mu\nu,\lambda\rho}(x,y)=\frac{g^{2}}{2N^{2}_{c}}<tr_{adj}\hat{F}_{\mu\nu}(x) \hat{\Phi}(x,y)\hat{F}_{\lambda \rho}(y)>,
\ee
	 here all dashed letters stand for operators in the adjoint representation.
At the next step we need to  find connection between 	BC and so called gluelump Green's functions .
Expanding $F_{\mu\nu}$ into
abelian (parentheses ) and nonabelian parts:\be
F_{\mu\nu}=(\partial_{\mu}A_{\nu}-\partial_{\nu}A_{\mu})-ig[A_{\mu},A_{\nu}],
\ee
we can write   BC as:
\be
D_{\mu\nu,\lambda\rho}(x,y)=D^{0}_{\mu\nu,\lambda\rho}(x,y)+D^{1}_{\mu\nu,\lambda\rho}(x,y)+D^{2}_{\mu\nu,\lambda\rho}(x,y),
\ee
where the number at the top of the letter D means power minus two of coupling constant g.  For $D^{0}(x,y)$ we obtain:
\be
D^{0}_{\mu\nu,\lambda\rho}(x,y)=\frac{g^{2}}{2N_{c}^{2}}( \frac{\partial}{\partial x_{\mu}} \frac{\partial}{\partial y_{\nu}} G^{1g}(x,y) +perm. )+\Delta^{0}_{\mu\nu,\lambda\rho},
\ee
where $\Delta^{0}_{\mu\nu,\lambda\rho}(x,y)$ contains contribution of higher field cumulants, which we systematically
discard. Here  $G^{1g}$ is one-gluelump Green's function:
\be
G^{1g}_{\mu\nu}(x,y)=<tr_{adj}\hat{A}_{\mu}(x)\hat{\Phi}_{adj}(x,y)\hat{A}_{\nu}(y)>, \label{1Gl}
\ee
 $tr_{adj}$ is a trace  over adjoint indices. As shown in Ref. \cite{16}, this term is connected with the functions $D^{E,H}_{1}$.From the physical point of view the Eq. (\ref{1Gl}) describes the gluon that is moving in the field of adjoint source (see Fig. \ref{Gl2} in Appendix A.5). Interaction between two objects in the adjoint representation is leading to formation of the string that according to Casimir scaling law found in the framework of FCM in \cite{SimShev} and supported by
 lattice data in \cite{Bali,Deldar} have a tension $\sigma_{adj}=\frac{C_{2}(adj)}{C_{2}(f)}\sigma_{f}=9/4 \sigma_{f}$, $C_{2}(adj),C_{2}(f)$ are Casimir operators for adjoint and fundamental representations. This hypothesis  give us a chance to calculate one-gluelump mass $M_{0}$. This mass governs  the nonconfinig part of the colour Coulomb's potential (we can calculate it from the correlator of Polyakov lines), and from our reasoning it is obvious that we can make the assumption that $M_{0} \sim \sqrt{\sigma_{adj}}$. From direct calculations we have  \cite{15}:
\be V_{1}(r,T)=-\frac{C_{2}(f)\alpha_{s}}{r}exp(-M_{0} r), M_{0} \simeq 2.06 \sqrt{\sigma_{s}}, rT\ll 1 \ee
 r-is a distance, T is  a temperature, $\alpha_{s}$ is a strong coupling constant. From comparison of the last equation with a simple  Debye potential \cite{Lebell} we can say that $M_{0}$ is playing the role of Debye mass.  \par
As for  $D^{2}_{\mu\nu,\lambda\rho}(x,y)$, it is of basic importance, since  ensures confinement via D(x-y) and
is expressed via two-gluon gluelump Green's function $G^{2g}(x,y)$.
The expression for $D^{2}_{\mu\nu,\lambda\rho}(x,y)$ reads as:
\be
D^{2}_{\mu\nu,\lambda\rho}(x,y)=-\frac{g^{4}}{2N^{2}_{c}}<tr_{adj}([A_{\mu}(x),A_{\nu}(x)]\hat{\Phi}(x,y)[A_{\lambda}(x),A_{\rho}(x)])> \label{35}.
\ee
We remind  that:
\be
[A_{i},A_{k}]=iA^{a}_{i}A^{b}_{k}f^{abc}T^{c}.
\ee
	Let's consider:
\be
G_{\mu\nu,\lambda\rho}(x,y)=tr_{adj}<f^{abc}f^{def}A^{a}_{\mu}(x)A^{b}_{\nu}(x)T^{c}\hat{\Phi}(x,y)A^{d}_{\lambda}(y)A^{e}_{\rho}(y)T^{f}>.\label{eqgluelump}
\ee
We
can fix in Eq. (\ref{eqgluelump}) color indices a, b; d, e and average  Green's function over all
fields $A^{h}_{\mu}$ with $h \ne a, b; d, e$.  This averaging will produce
the white string (of triangle shape at any given moment), and hence it will
ensure terms ($\delta_{ad} \delta_{be}$+ permutations). As a result we can represent $G_{\mu\nu,\lambda\rho}(x,y)$ in
the form:
\be
G_{\mu\nu,\lambda\rho}(x,y)=N^{2}_{c}(N^{2}_{c}-1)(\delta_{\mu\lambda}\delta_{\nu\rho}-\delta_{\mu\rho}\delta_{\nu\lambda})G^{2gl}(x,y), \label{38}
\ee
where $G^{2gl}(x,y)$ is the Green's function of the two-gluon gluelump.\par
Comparison of Eqs. (\ref{eqcorr1}), (\ref{35}) and (\ref{38}) immediately yields the following
expression for D(x-y):
\be
D(x-y)=\frac{g^{4}(N^{2}_{c}-1)}{2}G^{2gl}(x,y) \label{gl}.
\ee
Both one- and two-gluon gluelump  functions can be written in
terms of path integrals \cite{16B} and finally expressed via eigenvalues and eigenfunctions of relativistic string Hamiltonian \cite{19B}.

\section{The spatial string tension in the FCM}
We need to clarify some important moments:
at temperatures above deconfinement, $T > T_{c}$, large spatial Wilson loops still comply with 
the area law. For pure gauge SU(3) Yang-Mills theory $T_{c}=270 MeV$. This behaviour of
spatial Wilson loops \cite{13} is the main well established nonperturbative phenomenon at $T > T_{c}$,
which is usually called "magnetic
" or "spatial"
confinement. Of course, it does not contradict true deconfinement of a static quark-antiquark pair \cite{13* , 19,20,21}, because spatial-time Wilson
loops indeed lose the exponential damping with the area for $T > T_{c}$. To illustrate this phenomenon  one  can  calculate    polarization operator $\Pi(x,y)$ in Yang-Mills    theory \cite{Antonov}. As a standard step we need to decompose  gluon field  in non-perturbative  $B^{a}_{\mu}$ and perturbative $a^{b}_{\mu}$  parts:
\be
A^{a}_{\mu}=B^{a}_{\mu}+a^{a}_{\mu}, \label{decomp}
\ee
usually field B is treated  as   external.
The expression for  scalar part of polarization operator reads as:
\be
<\Pi(x,y)>_{B}=<tr (D^{2}[B]_{xy})^{-1}(D^{2}[B]_{yx})^{-1}>_{B} \label{Pol},
\ee
where   \be (D_{\mu}[B]a_{\nu})^{c}=\partial_{\mu}a_{\nu}^{c}+gf^{cde}B^{d}_{\mu}a^{e}_{\nu}. \ee

In  Eq. (\ref{Pol}) averaging is over $B^{a}_{\mu}$. At $T>T_{c}, T<<\sigma_{s} R, (\sigma_{s}=\sigma_{H})$  one observes  dramatic difference for spatial and time-like domains. For the first one  ($x_{0}-y_{0}=0,\vb{x}-\vb{y}=\vb{R}$)  one obtains at large distance:
\be
\Pi(0,R) \approx \frac{\sigma_{s}}{R^{2}}exp(-C\sqrt{\sigma_{s}}R), \sigma_{s}R^{2}\gg 1, C\approx 2\sqrt{2}, \label{Longdist}
\ee
and at small distance  polarization operator   tends to non-interacting case:
\be
\Pi(0,R) \to (\frac{1}{4\pi^{2} R^{2}})^{2} ,\sigma_{s}R^{2}\ll 1.
\ee
  For a time-like domain at zero temperature we have $\sigma_{E}=\sigma_{H}$, but at $T>T_{c}$ $\sigma_{E}=0$ as a manifestation of the confinement-deconfinement phase transition.  \par
	With these examples in mind we can focus on calculations of two-gluelump Green's function at non-zero temperature. The Eq. (\ref{eqgluelump}) describes two gluons moving in the field of adjoint source and interacting nonperturbatively (via $\sigma_{E,s}$) with  it  and between themselves.

The spatial string tension is proportional to the integral of  two-gluon  gluelump Green's function in the $3d$ space, where one of three space coordinates can be treated  as an evolution parameter ("`the Euclidean time"). Using the technic, developed in Ref. \cite{10,11,12}  we can write $G^{(2g)}_{4d} (z) =  G_{4d}^{(g)}\otimes G_{4d}^{(g)}$. We  neglect  the spin interactions in the first approximation. Concerning  $G_{4d}^{(g)}$
we have\cite{12,12*}:
\be
(-D^2)^{-1}_{xy}=\left\lan x\left|\int^{\infty}_0 dt
e^{tD^2(B)}\right|y\right\ran=
\int^{\infty}_0dt(Dz)^{w}_{xy}e^{-K}\Phi(x,y), \label{6}
\ee where
\be
K=\frac{1}{4}\int^s_0d\tau \left(\frac{dz_\mu}{d\tau}
\right)^2, ~~\Phi(x,y)=P \exp ig\int^x_yB_{\mu}dz_{\mu},\label{7}
\ee and a winding path measure is \be (Dz)^w_{xy}=\lim_{N\to
\infty}\prod^{N}_{m=1}\frac{d^4\zeta(m)}{(4\pi\varepsilon)^2}
\sum^{+\infty}_{n=-\infty}
\int\frac{d^4p}{(2\pi)^4}e^{ip(\sum\zeta(m)-(x-y)-n\beta\delta_{\mu
4})}. \label{8} \ee

The important point for the resulting $T$ dependence of the string tension is the integration in the gluon propagator $G^{(g)}_{4d}$  over the 4-th direction in Eq. (\ref{6}) with the exponent $K_4 = \frac14 \int^s_0 d\tau \left( \frac{d z_4}{d\tau}\right)^2,$ which gives for the spatial string tension  with $x_4=y_4$, and for the temporal string tension with the nonzero $x_4-y_4$  completely different behaviour, namely for the $\sigma_s$ case:
\be
J_4 \equiv \int (Dz_4)_{x_4x_4} e^{-K_4} = \sum^{+\infty}_{n=-\infty} \frac{1}{2 \sqrt{\pi s}} e^{-\frac{(n\beta)^2}{4s}} \label{imp}.
\label{9}
\ee

One can notice that the sum in the  Eq. (\ref{9}) is a known function:
\be
\sum^{+\infty}_{n=-\infty}e^{-\frac{n^2}{4sT^2}}\equiv\vartheta_3(q),~~~q=e^{-\frac{1}{4sT^2}},
\label{10a}
\ee
where the function $\vartheta_3(q)$ is defined as:
\be
\vartheta_3(q)=\sum^{+\infty}_{n=-\infty}q^{n^2}=1+2q+2q^4+O(q^9)
\label{10b},
\ee
and thus, the first term in this expression is connected with the vacuum contribution.
Then starting from low temperature there is an expansion:
$$
J_4=\frac{1}{2\sqrt{\pi s}}\sum^{+\infty}_{n=-\infty}e^{-\frac{n^2}{4sT^2}}
\equiv\frac{1}{2\sqrt{\pi s}}\vartheta_3(e^{-\frac{1}{4sT^2}})
$$
\be
=\frac{1}{2\sqrt{\pi s}}(1+2e^{-\frac{1}{4sT^2}}+O(e^{-\frac{1}{sT^2}}))
\label{10c}.
\ee
To find the asymptotics at high T one can use the relation:
\be
\sum^{+\infty}_{n=-\infty}e^{-\frac{\beta^2n^2}{4s}}=
\frac{2\sqrt{\pi s}}{\beta} \sum^{+\infty}_{n=-\infty} e^{-\frac{4\pi^2n^2}{\beta^2}s}
\label{10d}.
\ee
As a result at large T one obtains an equality:
$$
J_4=T\sum^{+\infty}_{n=-\infty}e^{-4\pi^2sT^2n^2}\equiv T\vartheta_3(e^{-4\pi^2sT^2})
$$
\be
=T(1+2e^{-4\pi^2sT^2}+O(e^{-16\pi^2sT^2}))
\label{10e}. \ee

Here we use the elliptic functions  $\vartheta_{3}(q)$  defined  in  Eq. (\ref{10b}).

\begin{figure}
\center{\includegraphics[width=0.7\linewidth]{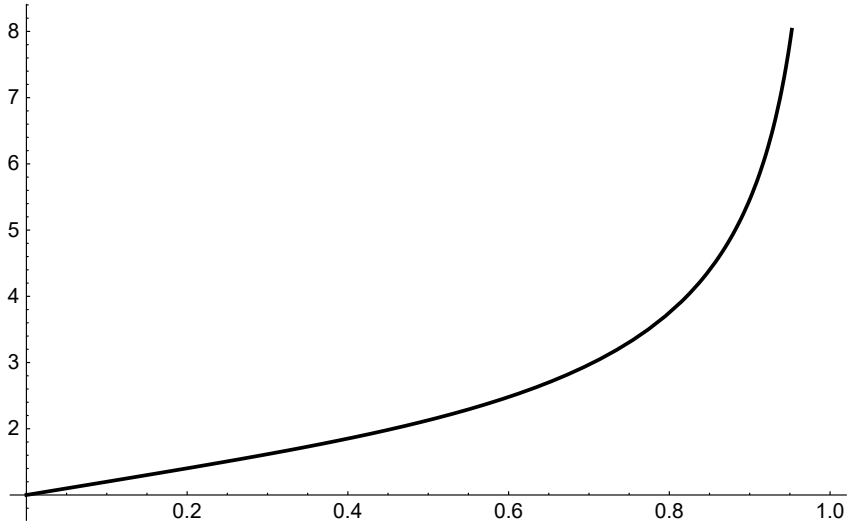}}
    \caption{ $\vartheta_{3}(q)$ as a function of $q$}
    \label{FiG1}
\end{figure}

Their behaviour as  function of $q$ is given in  Fig.\ref{FiG1}.
As a result we can use $J_{4}$ at an arbitrary T in the form:
\be
J_4(s,T)\equiv\frac{1}{2\sqrt{\pi s}}\vartheta_3(e^{-\frac{1}{4sT^2}})
\label{10f} .
\ee
In this way starting from  low T one obtains an exact expression for $J_4(T)$ valid in the whole range of $T$.
One could approximate this behaviour  as a sum of linear and constant terms implying a soft transition from
$T=0$ case to the linear in $T$ behaviour however this approximation fails numerically  and actually one
observes a sharp transition at some intermediate point $T^*$ from the regime $T=0$ to the large $T$ behaviour
as  given by the Eq. (\ref{10f}). For simplification we define:
\be
\vartheta_3(e^{-\frac{1}{sT^2}})= f(\sqrt{s}T).
\label{11}
\ee
At this point we turn to the general form of the field correlator $D^H(z)$ with the aim to express the string tension via the
factors $f(x)$. Rewriting the Eq. (\ref{gl}) with index "H"  we have:
\be
D^H(z) = \frac{g^{4}(T)(N^2_c-1) }{2} \lan G^{(2g)} (z,T) \ran,
\label{12}
\ee
where $G^{(2g)}(z,T)$ is the two-gluelump Green's function. In the path integral representation we can write it (see Appendix A2 for details) as:
\be
G^{(2g)}(z,T)= \frac{z}{8\pi} \int \frac{d\omega_1}{\omega_1^{3/2}} \frac{d\omega_2}{\omega_2^{3/2}}\int{} D^{2}r_1 D^{2}r_2
\exp{(-K_1-K_2-V(\ver_1,\ver_2)z)}I(z,T,\omega_{1},\omega_{2}), \nonumber\ee
\be I(z,T,\omega_{1},\omega_{2})=f(\sqrt{z/2\omega_1}T)f(\sqrt{z/2\omega_2}T) \label{13}. \ee
As a result we obtain $\sigma_s(T)$ in the following form:
\be
\sigma_s(T)= \frac{g^{4}(T)(N_c^2-1)}{4}\int d^2z z/(8\pi) \int d\omega_1 d\omega_2 (\omega_1\omega_2)^{-3/2}
\ee
\be
\times \sum_{n=0,1,} |\psi_n(0,0)|^2 \exp(-M_n(\omega_1,\omega_2)z) f(\sqrt{z/2\omega_1}T)f(\sqrt{z/2\omega_2}T)
\label{14}.
\ee
One can see in Eq. (\ref{14}) the only T-dependent factors $g^4(T)$ and $f(\sqrt{z/2\omega_1}T)$ which define the dependence of $\sigma_s(T)$. Therefore one can write $\sigma_{s}(T)$ (denoting the z- and $\omega$-integration in Eq. (\ref{14}) with the average sign
$<...>$) in the following form:
\be
\sigma_s(T)={\rm const} g^4(T) <f^2(\sqrt{z/(2\omega)T})>={\rm const} g^4(T) f^2(\sqrt{\overline{ \rm z/2\omega}} T), \ee
\be
\sigma_{s}(T)= {\rm const} g^4(T) f^2(\overline{\rm w}T),
 \label{16} \ee

  where we have denoted the average values of $\sqrt{\overline{\rm z/2\omega}}$ (obtained as a result of integration over the T-independent region of parameters with the T-independent kernel. We have also taken into account the symmetries of the Hamiltonian $H(\omega_{1},\omega_{2})$ with respect to permutation of $\omega_{1} $ and $\omega_{2}$ ) as $\overline{\rm w}= \rho/T_c$ and both $\rho,T_c$ are fixed parameters.
  The appearance of $g^4(T)$ which is decreasing with $T$ as $(\ln T)^{-2}$ defines the $T$
  dependence of $\sigma_s(T)$ to be lower than $T^2$, thus confirming the behaviour of $\sigma_s(T)$
  in the lattice data of \cite{8}, where the data were fitted
   as $\sigma_s(T)= {\rm const} g^4(T) T^2$ . However this fit fails for $T<2T_c$
   claiming the necessity of another factor in Eq. (\ref{16}).
   Correspondingly we are writing the resulting equation for the $\sigma_s(T)$ denoting the average value of
   $\sqrt{z/(2\omega)}T$ as $\rho T/T_c$.\par

In the next sections we try  to test our arguments and to demonstrate that this new form with the well-defined factor $f(\overline{\rm w}T)$
 describes the whole region of $T>T_c$ with  good accuracy.

\section{General expression for the spatial string tension vs lattice data}
For approve our predictions we need to find the parameter $\rho$ that describes the all data from the lattice simulations.  For  $f(\rho T/Tc)$ we have:
\be
 F(T/Tc)=f(\rho T/T_c)= \vartheta_3(e^{-\frac{T_c^2}{\rho^{2} T^2}})
\label{17}. \ee
 The numerical analysis of the data \cite{8} allows to
reproduce well the data with the Eq. (\ref{16}), derived in the previous  section 
 \be
\sigma_s(T)=\sigma_s(T_c) \frac{g^4(T) F^2(T/T_c)}{g^4(T_c)F^2(1)}
\label{18}. \ee

The comparison with  the lattice data of \cite{8} for Eq. (\ref{18}) is shown in Fig.2 , and the  expression for $g^4(T)$ is given  in the Appendix 1 and the value of the $\rho$-parameter $\rho=1/\sqrt{3.2}$. The Fig.2 demonstrates a good agreement between the lattice data and  Eq. (\ref{18}),
including the region $ T< 2.5 T_c$ where the lattice fit $T^2 g^4(T)$ in \cite{8} starts to disagree with numerical data.

\begin{figure}
\center{\includegraphics[width=0.9\linewidth]{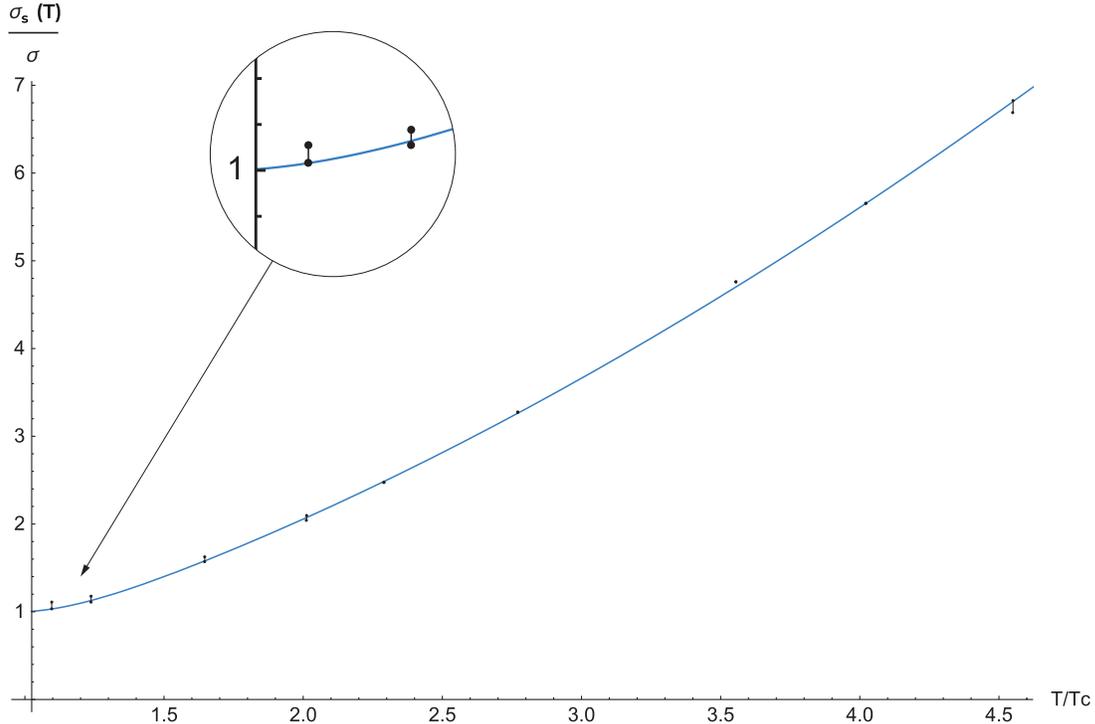}}
    \caption{Spatial string tension $\sigma_s (T)/\sigma$ for SU(3)
gauge theory as function of $T/T_c$.  The lattice data with errors are from Ref.\cite{8}. $T_{c}$=270 MeV}
    \label{FiG2}
\end{figure}
  The points on Fig. \ref{FiG2} as fuctions of $T/T_{c}$ completely coincide with the points from   Fig.3 in Ref.\cite{10*}. This fact means that at   sufficiently high temperatures  we numerically reproduce  the Eq.(\ref{3}). We also  show   the dependence  of running  coupling as function of  temperature (see Fig.\ref{Rel1} in the Appendix A1). As for the  dependence of $\frac{\vartheta_3(e^{-\frac{T_c^2}{\rho^{2} T^2}})}{T^{2}},\rho=1/\sqrt{3.2}$ it is equal to one with accuracy of two percent for temperatures higher than  1.3$T_{c}$ .  From these data we  also can  understand  that in the absence of  running coupling  the  string tension at high T is proportional to $T^{2}$.  Thus  one can conclude that running coupling has a small effect in comparison with $\vartheta^{2}_{3}$ . From all this facts we can say that temperature of dimensional reduction should be $T_{d}\geq 1.3 Tc$. That doesn't contradict neither the lattice data no earlier FCM predictions \cite{10*}.

\section{Discussion of results and Conclusions}

The main purpose of our work is the construction of the detailed mechanism of the spatial confinement
in the whole region of the temperature from $T=T_c$ to asymptotically large temperatures. The previous analysis in Ref. \cite{14}
has shown that the qualitative behaviour near $T=T_c$ can be continuosly connected with the asymptotic behaviour of the $\sigma_s(T)$. Due to complications of analytic calculations we tried to find the form of dependence of spatial string tension from the main QCD  parameters.  And  for this dependence  we   found the behaviour that well describes  the lattice data.    As  can be seen in  Fig.2 our resulting curve for the spatial string tension is in a good agreement with the
accurate lattice data in the whole measured region $T_c< T < 5 T_c$. We have exploited the coupling constant depending on the temperature $T$ given in the Appendix A1, which has also allowed the authors of \cite{8} to get agreement
with their data in the asymptotic region.  Also to describe  the region of smaller $T$ we have used the formalism
of the elliptic  functions $\vartheta_3(z)$ which describe well the sharp transition of the gluon propagator in the two-gluelump Green's function from the constant to the linear behaviour.It should be emphasized that the formalism presented in the paper is standard for the temperature dependence of any Green's functions developing in the spacial or time-like continuum with inclusion of interaction via the
field correlators \cite{10,11}.
 In particular the same spatial string tension appears in the expression for the  screening mass (we have called it   "Debye mass" $m_D(T)= 2.06 \sqrt{\sigma_s(T)}$ in  \cite{14,15}) characterizing the spatially oriented parts of the area of the Wilson loops. The inclusion of the temperature via the Matsubara-type formalism with the T-dependent factors $I(x_4-y_4,T)= exp(-ip_4(x_4-y_4 -n/T))$ in the gluon Green's functions is shown in the  Eq. (\ref{8}). For the space-like correlators with $x_4=y_4$  the use of the Poisson summation formula
$1/(2\pi) \sum_n \exp(ip_4 n\beta)= \sum_k \delta(p-4 \beta -2\pi k), \beta=1/T$, brings about an additional factor of $T$. This finally leads to the $T^2$ dependence of the leading term in the $\sigma_s(T)$ as in the  Eq. (\ref{3}).
  On the contrary for the time-like Green's functions with the nonzero $x_4-y_4$
the $T$ dependence is dictated by the corresponding mass parameters and  for  $\sigma_E(T)$ the situation is even more dramatic since it drops to zero (deconfinement) at $T=T_c$ approximately as $(1-(T/T_c)^4)^{1/2}$ \cite{22}. One can wonder why these two phenomena - spatial (colormagnetic) confinement and  colorelectric confinement are so different and hence disconnected and as follows from the lattice data (see Fig 9,10 in Ref. \cite{5}) the CE gluon condensate $<G_2^E (T)>$ and the CM condensate $<G_2^M (T)>$ being equal at $T=0$ behave also in a similarly different manner with  growing $T$?
The answer  lies in the different active regions of these phenomena - the space-like continuum for CM and
the time-like continuum for  CE confinement which have a little dynamical intersection as space-like
and time-like surfaces, which is evident in the FCM  and is an additional argument in favor of its selfconsistency. As it is, we have found   good agreement of our FCM approach for CM string tension with lattice data \cite{8} in this paper as well as good agreement of all our CE calculations with the corresponding lattice  and experimental data \cite{2,3,4,5,6,10,11,12,12*} including the latest CE calculations of the deconfining process  \cite{22}. Turning back to the CM physics it was found within
our approach that an
even more important  role of the spatial string tension may be  in the high $T$ thermodynamics where
in the framework of FCM it provides the basic nonperturbative contribution to the pressure and other observables, see e.g.  Ref.\cite{18}, in good agreement with the lattice data and solving as in Ref. \cite{12 } the old "Linde problem" which precludes pure perturbative thermodynamic calculations of interacting systems at large temperature. Another interesting development of this method is the dynamical theory of  QCD  systems
in the external magnetic field where the FCM yields all results in  good agreement with lattice data without any parameters, see e.g. \cite{ 23,24 }. In this way the FCM plays an important role in the
development of the present QCD theory.

\section{Acknowledgments}
The work of Khaidukov.Z.V was supported by the Russian Science Foundation 21-12-00237.
\section*{ Appendix A1. Two-loop expression for $g^{-2} (t)$}

\begin{figure}
\center{\includegraphics[width=0.7\linewidth]{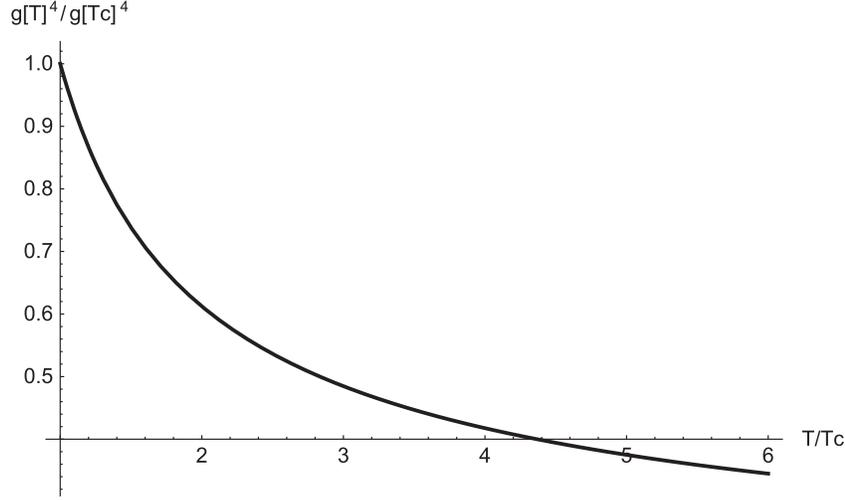}}
    \caption{Behaviour of $\frac{g(T)^{4}}{g(T_{c})^4}$ as a function of $\frac{T}{T_{c}}$. }
    \label{Rel1}
\end{figure}

 \setcounter{equation}{0} \def\theequation{A1.\arabic{equation}}
From the point of view of  FCM  the running coupling enters naturally in the formalism   qualitatively in the same way as in the standard theory  \cite{25}.
 One can
decompose the non-Abelian gauge field into the low- and the
high-energy parts: $A_{\mu}=B_{\mu}+a_{\mu}$.  At this step one can  treat $B_{\mu}$ as an external gauge field and at after that integrate out perturbative fields in the path integral. This procedure in the UV domain produces the running coupling constant and it can be used for calculation of  beta-function for example in three loops order   \cite{CR}. As a result one can use the two-loop beta-function calculations on the lattice in SU(3) gluodynamics \cite{8} where the  expression has the standard form as a function of $t= \frac{T}{T_c}$ :
\be
g^{-2}(t)= c_0 \ln \frac{t}{L_{\sigma}} + c_1 \ln\left(2 \ln\frac{t}{L_{\sigma}}\right),
\label{A1.1}
\ee
where
\be
c_0= \frac{11}{8\pi^2}, c_1= \frac{51}{88\pi^2}.
\label{A1.2}
\ee
Here $L_\sigma= \frac{\Lambda_\sigma}{T_c}= 0.104 \pm 0.009$ as in Refs. \cite{8,9}. All other parameters that we used for the calculations involving $g(t)$ are the same as in  Ref. \cite{8}. We also show  behaviour of $\frac{g(T)^{4}}{g(T_{c})^4}$ as a function of $\frac{T}{T_{c}}$ on Fig.\ref{Rel1}.

\section*{Appendix A2. Gluelump Green's function}

 \setcounter{equation}{0} \def\theequation{A2.\arabic{equation}}
There are two different ways to calculate two-gluelump Green's function and thus to obtain the spatial string tension from  Eqs. (\ref{gl}), (\ref{2}).   Two- gluelump Green's function  at non-zero temperature reads as:
\be
G^{2g}(x-y,T) =\int^{\infty}_{0}{ds}\int^{\infty}_{0}{d\bar{s}}\int{D^{w}z_{4}D^{w}\bar{z}_{4}}\int{D^{3}zD^{3}\bar{z}}exp(-S)<W(C_{z\bar{z}})> \label{glueA},
\ee
\be
S=\frac{1}{4}\int^{s}_{0}{d\tau}(\frac{d z_{\mu}}{d\tau})^{2}+\frac{1}{4}\int^{\bar{s}}_{0}{d\bar{\tau}}(\frac{d \bar{z}_{\mu}}{d \bar{\tau}})^{2},
\ee
x-y is a distance in four-space between beginning and  ending points.
The Wilson loop $W(C_{z\bar{z}})$  is averaged  over gluon fields along the paths z, $\bar{z}$ in the field of a static adjoint source with spatial coordinate $\vb{r}=(0,0,0)$  that  moves along z-axis entirely. This procedure  leads to formation of strings between gluons themselves and glouns and the  source. The  distance of movement  along z-axes is t. This interval must be large enough to avoid contribution of gluon fields fluctuations. With the help of   Eq. (\ref{imp}) one can integrate out fourth component of Eq. (\ref{glueA}) and obtain:
\be
G^{2g}(x-y,T) =\int^{\infty}_{0}{ds}\int^{\infty}_{0}{d\bar{s}}\int{D^{3}zD^{3}\bar{z}}exp(-S_{spatial})J_{s,\bar{s}}(s,\bar{s},T),
\ee
\be
J_{s,\bar{s}}(s,\bar{s},T)=\frac{1}{4\pi \sqrt{s\bar{s}}}\vartheta_3(e^{-\frac{1}{4sT^2}})\vartheta_3(e^{-\frac{1}{4\bar{s}T^2}}),
\ee
$S_{spatial}$ - is a  rest  part of the action without fourth component  containing "spatial" Wilson factor. Changing $s=\frac{t}{2\omega_{1}},\bar{s}=\frac{t}{2\omega_{2}}$  and Eq. (\ref{10f}) with third coordinate as "Euclidean time"\footnote{Such a choice  seems a bit surprising  but it  is  a price  for possibility to include finite temperatures in the formalism \cite{14}. That also means that   gluons  are moving  in the field of the static adjoint  source, that is evolving entirely along the z-axis.   } we can rewrite the last equation in the following form:
\be
G^{2g}(t,T)=t^{2}\int^{\infty}_{0}{\frac{d\omega_{1}}{2\omega_{1}^{2}}}\int^{\infty}_{0}{\frac{d\omega_{2}}{2\omega_{2}^{2}}}\int{D^{3}zD^{3}\bar{z}}exp(-S_{spatial})J(t,\omega_1 ,\omega_{2}),
\ee
\be
J(t,\omega_1 ,\omega_{2})=\frac{2\sqrt{\omega_{1}\omega_{2}}}{4\pi t }\vartheta_3(e^{-\frac{\omega_{1}}{2tT^2}})\vartheta_3(e^{-\frac{\omega_{2}}{2tT^2}}),
\ee
 and according to \cite{Simo} we can average  Wilson line over the " euclidean time direction"'.
 Thus we   obtain:
\be
G^{2g}(t,T) = \frac{t}{8\pi} \int^\infty_0 \frac{d\omega_1}{\omega_1^{3/2}} \int^\infty_0
\frac{d\omega_2}{\omega_2^{3/2}}\int{ (D^2z_1)_{xy}(D^2\bar{z}_2)_{xy}}
e^{-\sum_{i=1,2}K_{i}(\omega_{i})-Vt}\vartheta_3(e^{-\frac{\omega_{1}}{2tT^2}})\vartheta_3(e^{-\frac{\omega_{2}}{2tT^2}}) \nonumber, \\ \label{13*}
\ee

here label xy means xy-plane. For $K_i(\omega_i),i,j=1,2$:
\be
K_i(\omega_i)= \int^{t}_{0}d\tau_{E}(\frac{\omega_{i}}{2}+\frac{\omega_{i}}{2}(\frac{dx_{j}}{d\tau_{E}})^{2}).
\ee
The potential of interaction between gluons themselves and gluons with adjoint source reads as:
\be
V(z,\bar{z})=\sigma_{f}(|\vb{z}|+|\vb{\bar{z}}|+|\vb{z}-\vb{\bar{z}}|), \label{pot}
\ee
 $\sigma_{f}$ - is string tension in the fundamental  representation.
We can construct  the three-body Hamiltonian in the exponent of the  Eq. (\ref{13*})  in the $2d$ spatial coordinates:
\be
H(\omega_1, \omega_2) = \frac{ \omega_1^2+ \vb{p}^2_1}{2\omega_1}+  \frac{
\omega_2^2+ \vb{p}^2_2}{2\omega_2}+ V(\vb{z}, \vb{\bar{z}}). \label{15a}
\ee
If we obtain the spectrum of Eq. (\ref{15a})
we can rewrite the Eq. (\ref{13*}) as follows (see \cite{12}):
 \be
 G^{(2g)}(t) = \frac{t}{8\pi} \int^\infty_0 \frac{d\omega_1}{\omega_1^{3/2}} \int^\infty_0
\frac{d\omega_2}{\omega_2^{3/2}} \sum^\infty_{n=0} |\psi_n (0,0)|^2 e^{-M_n
(\omega_1,\omega_2) t}.\label{16a}
\ee
Here $\Psi_n(0,0)\equiv \Psi_n (\vez_1,\vez_2)|_{\vez_1=\vez_2=0}$, and $M_n$  is the eigenvalue of $H(\omega_1,
\omega_2)$. The latter was studied in  Ref.\cite{6} in three spatial coordinates. For our purpose here we only mention that the integral  in Eq. (\ref{14})  is  dimensionless.
One can note that procedure described above implies possibility of representation  of
 $4d$ gluon propagator at sufficiently large temperatures in the form:
\be
G_{4d}^{(g)}(z,T) = T G_{3d}^{(g)}(z)+K_{3d}(z).
\ee

\section*{Appendix A3. Diagonalization of  gluelump Hamiltonian}
 \setcounter{equation}{0} \def\theequation{A3.\arabic{equation}}
We can argue that  the Hamiltonian in Eq. (\ref{15a}) describes two non-interacting oscillators. Using standard technique we can rewrite it in the form:
 \be
 H= \frac{\omega^2_1 + \vep^2_1}{2\omega_1} + \frac{\omega^2_2 + \vep^2_2}{2\omega_2} + \frac{\sigma_{f}^2\vez^{2}}
{2 \nu_{1}} + \frac{\sigma_{f}^2 \bar{\vez}^2}{2 \nu_{2}} + \frac{\sigma_{f}^2 (\vez-\bar{\vez})^2}{2 \nu_{3}} + \frac{\nu_{1}+\nu_{2}+\nu_{3}}{2}.
\label{A3.1}
\ee
Canonical transformation of coordinates reads as:
\be
\vb{z}=\vb{x}+a \vb{y}, \bar{\vb{z}}=b\vb{x}+ \vb{y} \label{eqc11},
\ee
 and for momentums:
\be
 \vb{p_{1}}=c \vec{\pi}_{x}+ d \vec{\pi}_{y},\vb{p_{2}}=e \vec{\pi}_{x}+ f \vec{\pi}_{y} \label{eqm11},
\ee
where $a,b,c,d$ - are some parameters that we need to calculate.
Momentum and coordinates obey the relations:
\be
[p_{1,i},z_{j}]=-i\delta_{ij},[p_{2,i},\bar{z}_{j}]=-i\delta_{ij},[p_{2,i},z_{j}]=0,[p_{1,i},\bar{z}_{j}]=0,
\ee
\be
[\pi_{\alpha,i},x_{j}]=-i\delta_{\alpha x}\delta_{ij},[\pi_{\alpha,i},y_{j}]=-i\delta_{\alpha y}\delta_{ij}.
\ee
To define coefficients  in Eqs. (\ref{eqc11}), (\ref{eqm11}) we can substitute them into commutation relations and in the Hamiltonian  Eq. (\ref{A3.1}). Requiring of   vanishing of the cross terms ($\vb{\pi_{x}}\vb{\pi_{y}}, \vb{x}\vb{y}$)  we obtain:
\be
\omega_{1}a+\omega_{2}b=0, \nu_{1}^{-1}a+\nu_{2}^{-1}b=(a-1)(b-1)\nu_{3}^{-1} \label{coff1}.
\ee
For coefficients in Eqs. (\ref{eqc11}), (\ref{eqm11}) we have:
\be
c+ad=1,bc+d=0 \label{coff2},
\ee
\be
e+af=0,be+f=1 \label{coff3},
\ee
  thus:
\be
c=\frac{1}{1-ab},d=\frac{b}{ab-1}, f=\frac{1}{1-ab},e=-\frac{a}{1-ab}.
\ee
And for the Hamiltonian we have:
\be
H=(\frac{c^{2}}{2\omega_{1}}+\frac{e^{2}}{2\omega_{2}})\vec{\pi_{x}^{2}}+(\frac{d^{2}}{2\omega_{1}}+\frac{f^{2}}{2\omega_{2}})\vec{\pi^{2}_{y}}+V(\vb{x},\vb{y}) \label{Hamm1},
\ee
\be
V(\vb{x},\vb{y})=\frac{\sigma_{f}}{2}(\vb{x}^{2}(\frac{1}{\nu_{1}}+\frac{b^2}{\nu_{2}}+\frac{(1-b)^{2}}{\nu_{3}})+\vb{y}^{2}(\frac{a^{2}}{\nu_{1}}+\frac{1}{\nu_{2}}+\frac{(1-a)^{2}}{\nu_{3}})).
\ee
This is the Hamiltonian that describes two non-interacting two-dimensional oscillators and thus we can easily obtain its spectrum.
Minimizing Eq. (\ref{Hamm1}) with respect to $\omega_{1},\omega_{2},\nu_{1},\nu_{2},\nu_{3}$ we will obtain gluelump spectrum in the next section.
\section*{Appendix A4. Numerical calculation of gluelump  eigenvalues  and eigenfunctions in the lowest approximation}
\setcounter{equation}{0} \def\theequation{A4.\arabic{equation}}
We calculate here  eigenvalues and eigenfunctions of the Hamiltonian
 in the  Eq. (\ref{15a}).  We start from the expression:
 \be
 H= \frac{\omega^2_1 + \vb{p}^2_1}{2\omega_1} + \frac{\omega^2_2 + \vep^2_2}{2\omega_2} + \frac{\sigma_{f}^2\vez^{2}}
{2 \nu_{1}} + \frac{\sigma_{f}^2 \bar{\vez}^2}{2 \nu_{2}} + \frac{\sigma_{f}^2 (\vez-\bar{\vez})^2}{2 \nu_{3}} + \frac{\nu_{1}+\nu_{2}+\nu_{3}}{2}.
\label{A4.1}
\ee
For simplicity we assume that $\omega_1 =\omega_2=\omega$, $\nu_1 = \nu_2 =\nu_3 =\nu$.  \footnote{We have done that just for simplicity but already in this approximation  we obtain the lowest two-gluelump mass  with accuracy of 10 percent's.}  We  calculate all quantities  in Eqs. (\ref{eqc11}), (\ref{eqm11})
and obtain mass of the  lowest two-gluelump:
\be
E=(\sqrt{3}+1)\frac{\sigma_{f}}{\sqrt{\omega \nu}}+\frac{3\nu}{2}+{\omega}.
\label{A4.2}
\ee
The conditions of minima yields the
final result with notation $\omega,\nu$ for the extremal values. At final step we obtain\footnote{The question about   difference of two and one-gluelumps  masses  is of essential interest and it is  connected with QCD vacuum properties. We don't want to discuss it in this paper and only  mention that  possible corrections to gluelump masses  is discussed for example in \cite{16}.  }:
\be
M\approx 5.53 \sqrt{\sigma_{f}}.
\ee
As for the eigenfunctions, they are given by the product of the eigenfunctions of two non-interacting two-dimensional oscillators.
That means that we can  obtain all eigenvalues and eigenfunctions that we need in Eq. (\ref{14}).

\section*{Appendix A5. Gluelumpls Green fuctions}
\setcounter{equation}{0} \def\theequation{A5.\arabic{equation}}
\begin{figure}[h!]
\center{\includegraphics[width=0.7\linewidth]{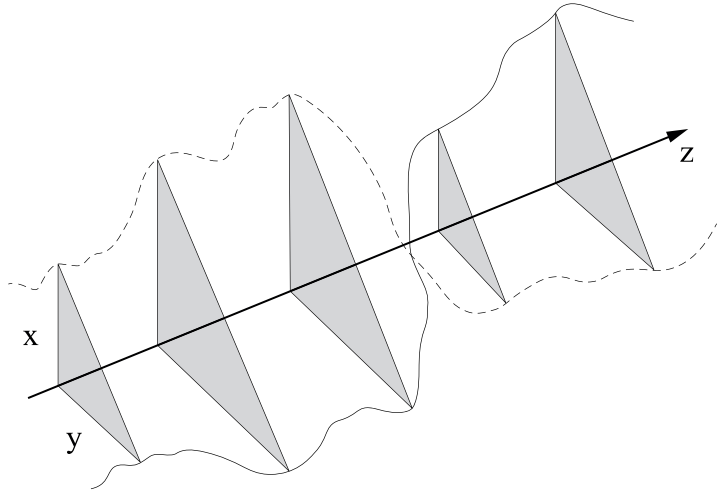}}
    \caption{ Two-gluelump Green function. Continuous and dashed lines are gluons trajectories. Bold straight line is trajectory of  adjoint source. Shaded domain  is x-y plane  that is perpendicular to z axes. Sides of shaded triangle are fundamental strings.  }
    \label{Gl2}
\end{figure}
\begin{figure}[h!]
\center{\includegraphics[width=0.7\linewidth]{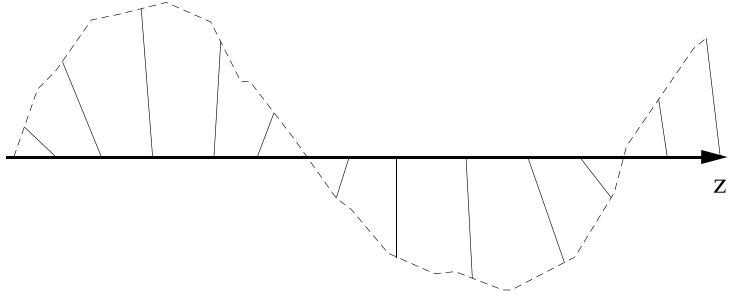}}
    \caption{ One-gluelump Green function.Bold straight line is trajectory of  adjoint source  that interacts with gluon (dashed line) through adjoint string). }
    \label{Gl1}
\end{figure}

\end{document}